\documentclass{aastex63}

\received{June 1, 2019}
\revised{January 10, 2019}
\accepted{\today}
\submitjournal{ApJ}

\shorttitle{Internetwork Magnetic Fields Seen in \ion{Fe}{1} 1564.8 nm}
\shortauthors{Hanaoka \& Sakurai}

\begin{document}

\title{Internetwork Magnetic Fields Seen in \ion{Fe}{1} 1564.8 nm Full-Disk Images}

\correspondingauthor{Yoichiro Hanaoka}
\email{yoichiro.hanaoka@nao.ac.jp}

\author[0000-0003-3964-1481]{Yoichiro Hanaoka}
\affiliation{National Astronomical Observatory of Japan \\
2-21-1 Osawa, Mitaka \\
Tokyo 181-8588, Japan}

\author[0000-0002-6019-5167]{Takashi Sakurai}
\affiliation{National Astronomical Observatory of Japan \\
2-21-1 Osawa, Mitaka \\
Tokyo 181-8588, Japan}

\begin{abstract}

We studied the properties of internetwork magnetic fields in the solar photosphere taking advantage of full-disk Stokes $V/I$ maps of the \ion{Fe}{1} 1564.8 nm line, which were obtained during 2010--2019. 
In contrast to most of the previous studies, we used data with moderate spatial and spectral resolutions. Nonetheless, we were able to distinguish the internetwork field components and the active region / network boundary components using large Zeeman splitting of the \ion{Fe}{1} 1564.8 nm line. Thus, our analysis provides a point of view quite different from that of the previous studies.
We analyzed the data statistically without ordinary inversions, yet we successfully derived some properties of internetwork fields; the internetwork is filled with small-scale magnetic fields, their strength is within the weak-field regime of the \ion{Fe}{1} 1564.8 nm line (300--400 G or less), and the internetwork fields are highly inclined. Although the results were obtained from the analysis performed from a different perspective, they are consistent with the majority of the previous findings. In addition, no notable variation in the properties of the internetwork fields was found during the period covering most of solar cycle 24. 

\end{abstract}


\section{Introduction} \label{sec:intro}

The solar surface is filled with magnetic field. Moreover, active regions and supergranulation network boundaries in the quiet Sun, which have strong magnetic fields, are its dominant components. However, magnetic fields in internetwork regions inside the networks are also an important component of the solar magnetic field in spite of their weak field strength. According to one recent estimation, the strength of the internetwork fields is approximately 130 G at the optical depth of unity \citep{2016A&A...593A..93D}, and it is presumed that their flux occupies 14 \% of that of the quiet Sun magnetic field \citep{2014ApJ...797...49G}. The fact that the weak magnetic field spreads over the internetwork regions was noted by \citet{1971IAUS...43...51L} and many studies have been conducted. In particular, both ground-based and space-borne sophisticated instruments have enabled advanced observations of the weak and small-scale elements of internetwork fields, which have contributed to the progress of the research of internetwork fields.

\citet{2019LRSP...16....1B} summarized such studies in their review paper. They presented widely accepted properties of internetwork fields as some ``agreed facts''; namely, internetwork fields are mostly weak (the order of hectogauss (hG)), highly inclined, and have large filling factors. However, the interpretations derived from various observations still show discrepancies. Internetwork fields are challenging targets because of their small scale and weak field strength, and the observations carried out with different techniques have not necessarily reached consistent conclusions. For instance, the results based on absorption lines in visible wavelengths and those based on infrared lines are sometimes inconsistent with one another.

The visible wavelength observations, which realize high spatial resolutions and high polarization sensitivities, are suitable to measure small-scale weak magnetic fields in internetwork regions.
On the other hand, polarization measurements of infrared lines at \ion{Fe}{1} 1.56 $\mu$m (1564.8 nm and 1565.3 nm) have also been a powerful tool for the study of internetwork fields since \citet{1995ApJ...446..421L}. In particular, the \ion{Fe}{1} 1564.8 nm line shows much larger Zeeman splitting than visible lines such as the \ion{Fe}{1} 630 nm line because of its large Land\'{e} factor $g$ and its long wavelength. The magnetic field of the order of kilogauss (kG) in the active regions or the network boundaries is beyond the ``weak-field'' regime of the \ion{Fe}{1} 1564.8 nm line, where the splitting of the line does not behave proportionally to the magnetic field strength but the degree of polarization does to a certain degree \citep[e.g.,][]{1992soti.book...71L}. Therefore, it is possible to distinguish the weak hG field of the internetwork and the kG field directly by using Zeeman splitting and without using inversions.

One of the most important topics on which the results derived from the visible line observations and those from the infrared observations sometimes show discrepancy is the inclination distribution of internetwork fields.
Results in which the horizontal components dominate in internetwork fields have often been deduced from visible wavelength observations. 
This was discovered for the first time by \citet{1996ApJ...460.1019L}, and confirmed by more recent observations with space-borne instruments (e.g., \citet{2007ApJ...670L..61O}, \citet{2008ApJ...672.1237L}, and \citet{2010ApJ...713.1310I} using the Solar Optical Telescope \citep[SOT;][]{2008SoPh..249..167T} of Hinode \citep{2007SoPh..243....3K}) and with ground-based instruments (e.g., \citet{2013A&A...555A.132S} using the Heliographic Telescope for Studying Magnetism and Solar Instabilities (THEMIS) of the Mt. Teide Observatory at Tenerife).
On the other hand, \citet{2003A&A...408.1115K}, \citet{2008A&A...479..229M}, \citet{2009A&A...502..969B}, and \citet{2016A&A...596A...5M} concluded that the highly inclined field cannot be seen in the observations of the \ion{Fe}{1} 1.56 $\mu$m lines, while some results based on the \ion{Fe}{1} 1.56 $\mu$m lines consistent with those from visible lines have also been published \citep{2006ApJ...646.1421D, 2008A&A...477..953M}. 

In this study, we attempted to shed light on this topic from a different perspective using full disk images of the \ion{Fe}{1} 1564.8 nm line.
As mentioned above, the observations of internetwork fields have mostly been undertaken with advanced large telescopes using a small field of view to pursue a high spatial resolution and a high polarization sensitivity. However, the data used here were taken with a moderate spatial resolution.
The solar group of the National Astronomical Observatory of Japan (NAOJ) has been carrying out a regular full-disk solar observation with a near-infrared spectropolarimeter \citep{2018PASJ...70...58S} installed in the observatory's Solar Flare Telescope (SFT; \citealp{1995PASJ...47...81S}; for its recent status, see \citealp{2020JSWSC..10..41H}) from 2010 to present. The full-Stokes data of the \ion{Fe}{1} 1564.8 nm along with He I 1083.0 nm and Si I 1082.7 nm are acquired during observation. The high polarimetric sensitivity was demonstrated by \citet{2017ApJ...851..130H}, who statistically analyzed linear polarizations observed in solar filaments in He I 1083.0 nm.
The spectral and spatial resolutions of the data used here are not as high as those in the previous studies using the \ion{Fe}{1} 1564.8 nm line. Therefore, the signal in a pixel is probably due to a mixture of unresolved internetwork field elements, and it is difficult to carry out genuine Stokes inversions as in the previous studies performed with high resolution data. However, we can examine the Stokes $V/I$ (degree of circular polarization) profiles, which are reliably measured even in lower-resolution observations but with high polarimetric sensitivities, statistically utilizing large amount of data accumulated for years \citep[as briefly reported by][]{2015IAUS..305...92H}. 

One of the benefits of the data analyzed here is that they are full-disk data showing the Stokes signals from the center to the limb simultaneously. This fact is important because the center-to-limb variation of the polarization signals provides a hint for the inclination distribution of the internetwork fields. 
The horizontal magnetic field presents no line-of-sight magnetic field component at the disk center, but its line-of-sight component increases toward the limb with the distance from the disk center, and the circular polarization will appear. If the horizontal magnetic field has a randomly distributed azimuth as discussed in the previous studies, it will be observed as fluctuating circular polarization signals. On the other hand, the vertical magnetic field, which is dominant in the kG components, shows the limbward decrease of the line-of-sight component.
\citet{2007ApJ...659L.177H} demonstrated the manifestation of internetwork fields in the full disk magnetograms taken with the Synoptic Optical Long-term Investigations of the Sun \citep[SOLIS;][]{2003SPIE.4853..194K} and the Global Oscillation Network Group \citep[GONG;][]{1988ESASP.286..203H} instruments using visible wavelength lines. They found a limbward increase of the line-of-sight magnetic field components, which is consistent with the highly inclined field. This result is also supported by the studies using high resolution observations of the center-to-limb variation by \citet{2008ApJ...672.1237L} and \citet{2013A&A...555A.132S}, and particularly by \citet{2017ApJ...835...14L}, who analyzed a large amount of linear polarization data (representing the transverse magnetic field component) obtained at various distances from the disk center with the Hinode SOT. They found an increase of the linear polarization components parallel to the limb with the distance from the disk center in the internetwork fields. As noted by \citet{2013A&A...555A.132S}, this is the behavior of the highly inclined field.
However, with 1.56 $\mu$m observations, \citet{1998A&A...331..771M} found the limbward {\it decrease} of the weak magnetic field signal, and \citet{2008A&A...479..229M} concluded that there is {\it no} center-to-limb variation in internetwork fields.

Another benefit of the data used here is that they cover most of solar cycle 24; therefore, we can study the solar cycle dependence of the polarization signals of the internetwork regions. 
Some researchers have studied the internetwork field data obtained at different phases in the solar activity. For example, \citet{2003A&A...411..615S}, \citet{2013A&A...555A..33B}, and \citet{2014PASJ...66S...4L} concluded that there is no cycle variation. On the other hand, other studies have shown the solar cycle dependence of the weak magnetic field on the solar surface \citep{2001A&A...378..627F, 2011ApJ...731...37J, 2015A&A...582A..95F}. However, \citet{2015A&A...582A..95F} inferred that the variation was caused by noise or contamination from active regions, and \citet{2001A&A...378..627F} and \citet{2011ApJ...731...37J} presumed that the measured weak magnetic field was not necessarily the internetwork fields. These previous studies used visible wavelength lines, but this study is the first to approach the solar cycle variation of the internetwork fields observed with the \ion{Fe}{1} 1564.8 nm line.

Thus, we conducted a study utilizing full-disk images to better understand the nature of internetwork fields from a perspective different from those of the previous studies. In this paper, an explanation of the data is presented in Section 2, results of the analysis are given in Section 3, and conclusions are summarized in Section 4.

\section{Observation and Data} \label{sec:obs}

\subsection{Instrument and Observation}

Because the infrared spectropolarimeter of the SFT is described in detail by \citet{2018PASJ...70...58S}, we briefly explain its specifications in relation to the \ion{Fe}{1} 1564.8 nm observation and data. It started the observation in 2010, and the data obtained during the period of 2010--2019 were used for this study. The telescope tube has a 15-cm objective lens, and a ``single-beam'' spectropolarimeter using only one of the orthogonal linear polarizations is placed downstream. The slit of the spectrograph, which is set along the celestial north-south direction, covers half of the solar diameter. The slit sweeps the northern and southern hemispheres separately to cover the full Sun. Samplings on each spectral image of the \ion{Fe}{1} 1564.8 nm line are approximately 2$''$ spatially and 6.3 or 6.7 pm (63 or 67 m\AA) spectrally. In 2015, the optical system was modified to accommodate two cameras for simultaneous observation of the two wavelength bands around \ion{He}{1} 1083.0 nm and \ion{Fe}{1} 1564.8 nm. This change introduced a reduction of the image size on the camera detector of approximately 7 \%. The spectral sampling changed from 6.3 to 6.7 pm, and the spatial sampling along the slit changed correspondingly. The step width of the slit scan was approximately 3$''$ at first, and it was changed to 2$''$ in 2011 July. Therefore, a scan have been provided a data cube covering the full disk with the spatial sampling of 2--3$''$ and the spectral sampling of 6.3 -- 6.7 pm. The true spatial resolution taking the diffraction limit and the seeing degradation into account is 3--4$''$, and the nominal spectral resolution determined by the slit and the grating is 6.7 pm. It takes about 3.5 s to obtain a data set at a single slit position, and to complete a full-Sun scan, approximately two hours is required. Weather permitting, 1--3 full-disk scans are performed daily. 

Currently we are using a rotating waveplate as the polarization modulator. The typical root-mean-square (RMS) noise level of the Stokes $V/I$ signal in a pixel of the detectors, namely the polarization sensitivity, is about 0.00016. Thanks to the high full-well capacity ($\sim10^6$ e$^-$) of InGaAs detectors of the cameras and the high efficiency of the polarization modulation ($\sim70$ \%), this low noise level is achieved by combining 192 polarization-modulated images taken at each slit position, whose total exposure time is 2 s. Until 2013, we used ferroelectric liquid crystals as the polarization modulator. Due to the low efficiency of the polarization modulation, about one tenth of the current value, the typical noise level was about 0.00046. However, even before the replacement of the polarization modulator, the noise level is low enough to enable the detection of the weak polarization signals in the internetwork regions.

\subsection{Preparation of the Data for the Analysis}

\begin{figure}
\plotone{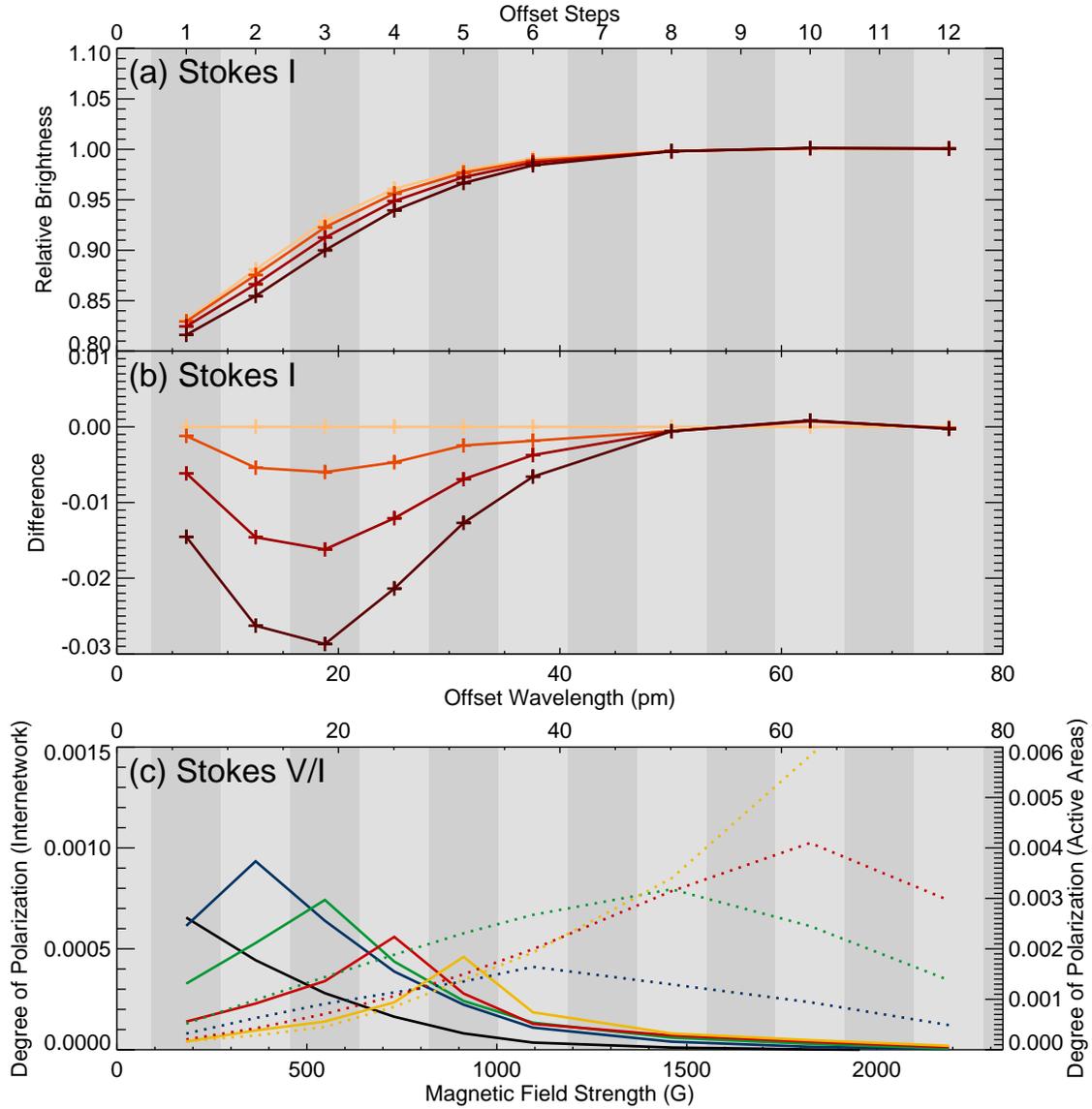}
\caption{Average Stokes $I$ and $V/I$ profiles of the \ion{Fe}{1} 1564.8 nm line obtained from the data acquired on 2014 May 10. The blue and red wings are folded, and the $x$-axis represents the offset from the line center. Panel (a) shows the Stokes $I$ profiles of the internetwork regions in the four zones shown in Figure \ref{fig:fig4} from the disk center area (light brown) toward the limb (dark brown).
Panel (b) shows the relative Stokes $I$ profiles with respect to the disk center area. In panel (c), the Stokes $V/I$ profiles that have their peak positions at various offsets from the line center are shown. Solid lines represent the average profiles of the internetwork pixels (plotted using the scale on the left), and dotted lines represent those of the active regions or network boundaries (plotted using the scale on the right). The magnetic field strengths, which give the Zeeman splitting of the corresponding wavelength offsets, are also provided in the $x$-axis. The gray stripes in the background represent the wavelength sampling steps.
  \label{fig:fig1}}
\end{figure}

We selected 99 sets of full-disk scan data of \ion{Fe}{1} 1564.8 nm for analysis from the daily data obtained from 2010--2019; or, roughly one data set per month. The selected data were all acquired under fairly stable weather conditions and well-kept focus adjustments. The Carrington longitudes of the central meridian of the selected data were roughly uniformly distributed.
Because we intended to carry out a statistical study utilizing a large amount of data, we focused only on the Stokes $V/I$ signals (degree of circular polarization derived from the Stokes $V$ and $I$ values measured in the same pixel) and examined the peak(s) in each Stokes $V/I$ profile. The positive and negative peaks in a Stokes $V/I$ profile represent the $\sigma$-components shifted from the line center by the Zeeman effect.

The Stokes data used in the analysis were prepared as follows.
Raw observed spectra were discretely sampled by the detector pixels; the sampling was not necessarily symmetric to the center of the \ion{Fe}{1} 1564.8 nm line.
Therefore, to analyze the individual Stokes profiles, we first calculated the Stokes signals at the wavelengths distributed symmetrically with respect to the line center using interpolation, keeping the wavelength sampling step (6.3 -- 6.7 pm) unchanged from the original. The wavelengths were $\pm$1, $\pm$2, $\pm$3, $\pm$4, $\pm$5, $\pm$6, $\pm$8, $\pm$10, and $\pm$12 steps from the line center.
This wavelength range covers approximately $\pm$6 -- 80 pm ($\pm$60--800 m\AA ) from the line center. In addition, it corresponds to the magnetic field strengths of up to approximately 2 kG in terms of the Zeeman splitting, because the displacement from the line center is expressed as $\Delta\lambda = 4.67\times 10^{-12}gB\lambda_0^2$ nm, where $g=3$ is the Land\'{e} factor, $B$ is the magnetic field strength in gauss, and $\lambda_0$ is the wavelength of the line in nm.
As mentioned above, the wavelength sampling step changed from 6.3 to 6.7 pm in 2015. Hence, hereafter we express the distance from the line center with the number of offset steps as well as the offset wavelengths.
For all spectral data, we fitted the bottom position of the \ion{Fe}{1} 1564.8 nm line with a parabolic curve and derived the line center position. This process was applied to the spectral data at each slit position separately; thus, the defined line center is free from the Doppler shift owing to the solar rotation.

Next, we folded the Stokes profiles in the blue and red wings into a single wing profile. For the Stokes $V/I$ profiles, which are antisymmetric to the line center, the signals in the red wing were subtracted from those in the blue wing. Figures \ref{fig:fig1}(a) and \ref{fig:fig1}(c) show some folded Stokes $I$ and $V/I$ wing profiles (discussion for Figure \ref{fig:fig1} will be presented later). In Figure \ref{fig:fig1}(c), the average Stokes $V/I$ profiles in the internetwork regions and the active regions / network boundaries, which peaked at each of 1, 2, 3, 4, 5, 6, 8, 10, and 12-steps offsets from the line center, are plotted. We defined the peak position of each folded Stokes $V/I$ profile as the wavelength offset where the absolute degree of circular polarization becomes the maximum. These profiles, which have different peak positions, are considered to represent different magnetic field strengths as shown in the $x$-axis of Figure \ref{fig:fig1}(c). However, in the weak-field regime, the peak Stokes $V/I$ is always located where the gradient of the Stokes $I$ profile takes the largest value. If we take the effective line width in our observation degraded by the spectral resolution into account and ignore the Doppler shift of the line, the wavelength of the maximum gradient is located at 2-steps from the line center, which approximately corresponds to 300--400 G in terms of the Zeeman splitting. Therefore, if the peak position of a Stokes $V/I$ profile is located at the offset of 2-steps, it is presumed that the magnetic field strength is 300--400 G or less. In this way, the peak position of each Stokes $V/I$ profile can be connected to a magnetic field strength without an inversion.

Thus, the prepared data were statistically analyzed mainly using the peak positions as an indicator of the magnetic field strength; the results are presented in the next section.

\section{Results} \label{sec:results}

\subsection{Properties of the Stokes $V/I$ Signals Seen in Full-Disk \ion{Fe}{1} 1564.8 nm Maps} \label{subsec:tables}

Here, we present the properties of the Stokes $V/I$ signals in the full-disk maps of the \ion{Fe}{1} 1564.8 nm line. First, we show typical example data obtained on 2014 May 10 (around the solar maximum) and 2019 August 10 (around the solar minimum). 

\subsubsection{Appearance of the Full-Disk Stokes $V/I$ Maps of \ion{Fe}{1} 1564.8 nm}

\begin{figure}
\plotone{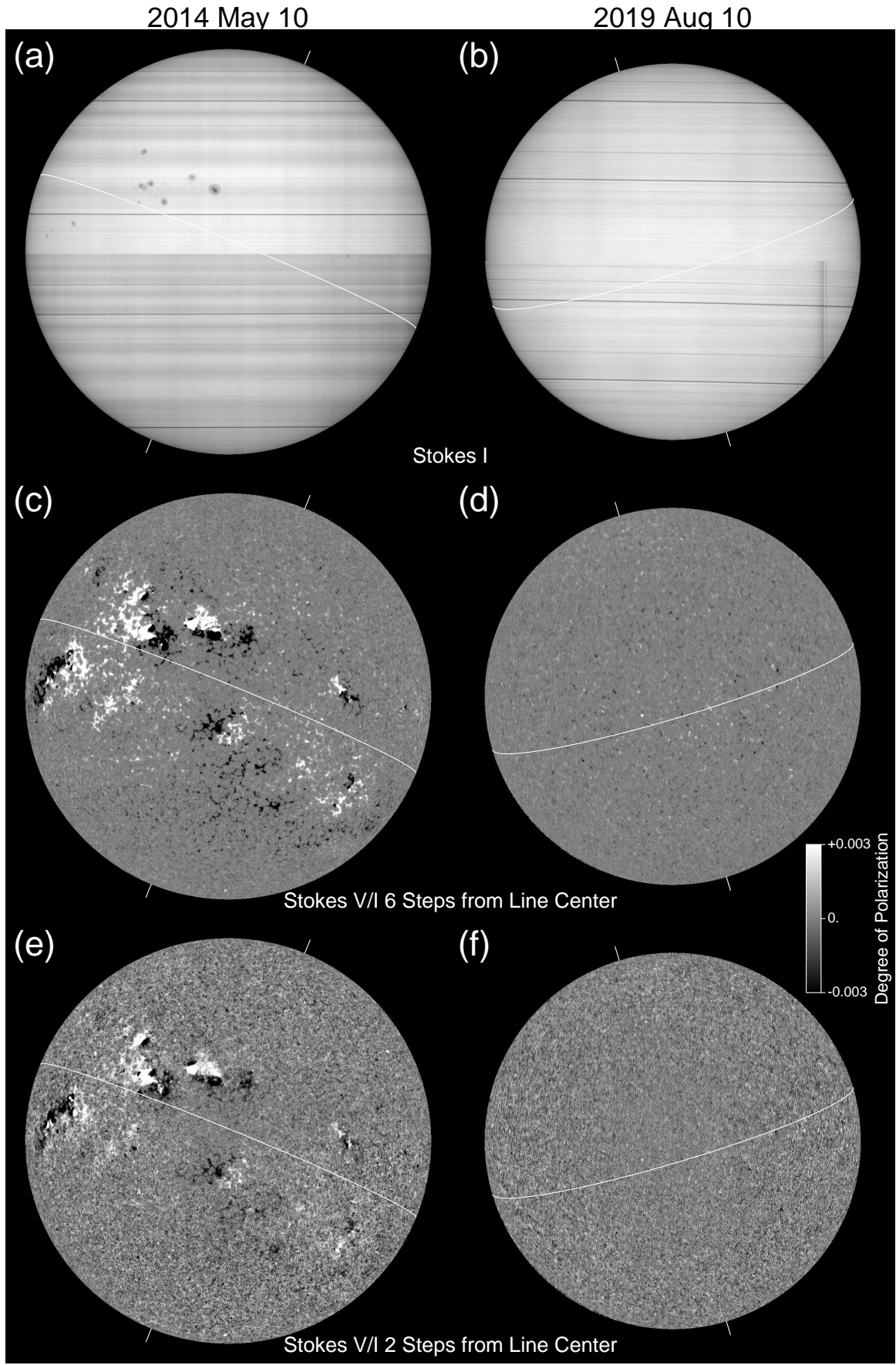}
\caption{Sample Stokes $I$ (panels (a) and (b)) and $V/I$ maps ((c)--(f)) of the \ion{Fe}{1} 1564.8 nm obtained on 2014 May 10 and 2019 August 10. The celestial north is to the top. Panels (c) and (d) are maps of the Stokes $V/I$ signals at the 6-steps offset ((c) 37.7 / (d) 40.3 pm) from the line center; panels (e) and (f) are those at the 2-steps offset ((e) 12.6 / (f) 13.4 pm). Regardless of the solar activity, the Stokes $V/I$ maps at the 6-steps offset and the 2-steps offset show remarkable differences, particularly in the internetwork regions.
  \label{fig:fig2}}
\end{figure}

Figure \ref{fig:fig2} shows the Stokes $I$ and Stokes $V/I$ maps of \ion{Fe}{1} 1564.8 nm on the two aforementioned dates. The Stokes $I$ maps shown in Figures \ref{fig:fig2}(a) and \ref{fig:fig2}(b) show that the number of sunspots on these two dates were quite different. As the spatial sampling was approximately 2$''$, each map is comprised of approximately 1000$\times$1000 pixels. For each pixel, the Stokes $V/I$ signals at the offsets of 1 --12 steps from the line center were calculated as mentioned above. In Figures \ref{fig:fig2}(c)-- \ref{fig:fig2}(f), the Stokes $V/I$ maps at the wavelength offsets of 6 and 2 steps are shown. The actual wavelength offsets are 37.7 and 12.6 pm for 2014 May 10 and 40.3 and 13.4 pm for 2019 August 10. 

The maps at 6-steps offset, shown in Figures \ref{fig:fig2}(c) and \ref{fig:fig2}(d), seem to be ordinary longitudinal magnetograms showing remarkable polarization signals in and around the active regions or network boundaries in the quiet Sun. On the other hand, the maps at 2-steps offset from the line center, Figures \ref{fig:fig2}(e) and \ref{fig:fig2}(f), are much different from those at 6-steps offset. The whole disk shows grainy appearance; it is filled with small scale positive and negative signals. They look as if they are random noise, but actually they fluctuate about five times more than the noise level mentioned in Section 2.1. Taking a closer look, we can find that the polarization signals are rather weak around the disk center, similarly to the finding by \citet{2007ApJ...659L.177H} in full-disk magnetograms. This is an opposite tendency to the signals seen in Figures \ref{fig:fig2}(c) and \ref{fig:fig2}(d), which show apparent limbward weakening. 
Contrary to the active regions and network boundaries seen in Figures \ref{fig:fig2}(c) and \ref{fig:fig2}(d), where the vertical magnetic fields dominate, the grainy signals are considered to represent highly inclined magnetic fields, as explained in Section 1. 

From Figure \ref{fig:fig2}, it is expected that the peak in Stokes $V/I$ profiles of the active regions or network boundaries are located far from the line center, while those of the grainy signals are located around the wavelength offset of 2 steps.
The wavelength offset of 6 steps or 37.7 / 40.3 pm corresponds to the Zeeman splitting of the magnetic field of 1.1 k / 1.2 kG as shown in Figure \ref{fig:fig1}(c), and therefore, it is natural that the so-called kG components, active regions and network boundaries in the quiet Sun, show remarkable polarizations in Figures \ref{fig:fig2}(c) and \ref{fig:fig2}(d).
On the other hand, the wavelength offset of 2 steps (12.6 / 13.4 pm) corresponds to the Zeeman splitting of 365 / 392 G. Therefore, the polarization signals, which are more remarkable in the Stokes $V/I$ maps at 2-steps offset than in those at 6 steps offset, are presumed to correspond to the magnetic field of the hG order.

From the above, the grainy signals seen in Figures \ref{fig:fig2}(e) and \ref{fig:fig2}(f), which, in principle, would correspond to highly inclined hG magnetic fields and are spread over the entire solar disk, can be interpreted as the internetwork magnetic fields. This means that the internetwork magnetic fields were captured in the full-disk Stokes $V/I$ maps of the \ion{Fe}{1} 1564.8 nm, even though the observations were made with a moderate spatial resolution of 3--4$''$. Next, we present the behavior of the peak positions of the Stokes $V/I$ signals.

\subsubsection{Distribution of the Offsets of the Peak Positions in the Stokes $V/I$ Profiles}

\begin{figure}
\plotone{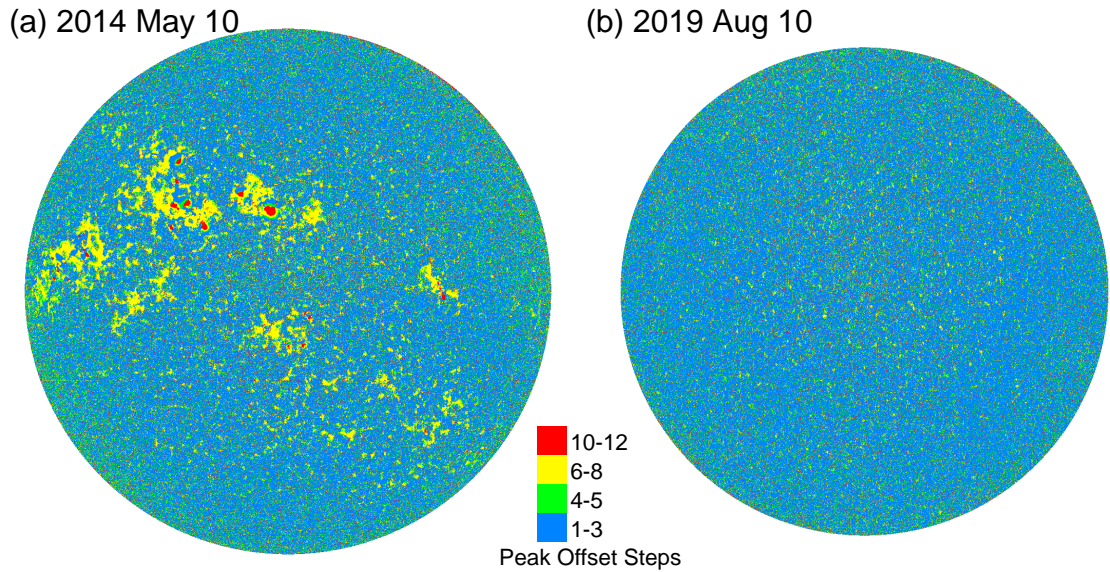}
\caption{Distribution of the offsets of the peak positions in the Stokes $V/I$ profiles from the center of the \ion{Fe}{1} 1564.8 nm line for the data on 2014 May 10 and 2019 August 10. The peak offsets are classified into four groups represented by different colors.
  \label{fig:fig3}}
\end{figure}

Figure \ref{fig:fig3} shows the spatial distribution of the offsets from the line center of the peak positions in the Stokes $V/I$ profiles of individual pixels in the full-disk maps.
Panels (a) and (b) show the data for 2014 May 10 and 2019 August 10; the corresponding Stokes $V/I$ maps are presented in Figures \ref{fig:fig2}. The wavelength offsets of the peak positions are classified by color in Figure \ref{fig:fig3}. Pixels with the offsets of 10 or 12 steps are shown in red, and those with the offsets of 6 or 8 steps are shown in yellow.
The offsets of 6--12 steps roughly correspond to the Zeeman splitting of the magnetic field strength of 1--2 kG. They are particularly conspicuous in Figure \ref{fig:fig3}(a), and they approximately coincide with the active regions or network boundaries in the quiet Sun, as shown in Figure \ref{fig:fig2}(c). In Figure \ref{fig:fig3}(b), we see yellow patches corresponding to the network boundaries, which are seen as white and black spots sparsely scattered over the Stokes $V/I$ map in Figure \ref{fig:fig2}(d).
By contrast, the pixels in blue, which have the peak offset within 3 steps from the line center, dominate the rest of the pixels and are spread over the entire disk both in Figures \ref{fig:fig3}(a) and \ref{fig:fig3}(b). The pixels with peak offsets of 4 or 5 steps are shown in green and are also scattered over the entire disk. Particularly in Figure \ref{fig:fig3}(b), which presents the data taken during low solar activity, most of the pixels are painted in blue or green. 
It is evident that the distribution of the offsets of the peak positions is generally consistent with that of the kG fields in the active regions or network boundaries and hG fields in the remaining areas shown in Figure \ref{fig:fig2}. 

On the other hand, we can find a certain number of red or yellow pixels scattered over the entire disk. Since they are mostly isolated with a low degree of polarization, they are considered to be the result of noise. If there is no magnetic field, or if mixed polarities of positive and negative components of the magnetic field cancel each other out in a pixel, the Stokes $V/I$ signal remains at zero or is very weak for an entire profile. In such cases, noise dominates the measured polarization signals, and the Stokes $V/I$ peak appears at any wavelength and is offset randomly. Such pixels may appear anywhere outside of the active regions or the network boundaries. 
There have been some reports in which the kG components are present in the internetwork magnetic fields; however, these have been hypothesized to be spurious signals produced through incomplete measurements or inversions \citep[see][and references therein]{2019LRSP...16....1B}. With regard to our data, the majority of such pixels have noise-dominant profiles.

\subsubsection{Defining the Internetwork Pixels in the Stokes $V/I$ Maps}

\begin{figure}
\plotone{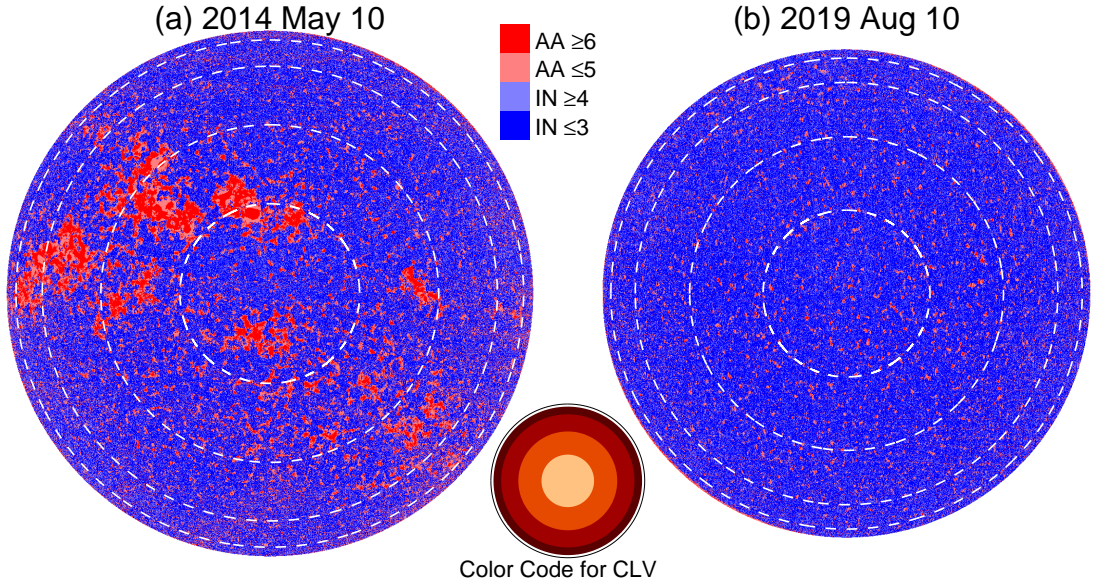}
\caption{Distribution of the active region and network boundary pixels (AA [active area]; red and light red, degrees of polarization exceeding three times the noise level) and that in the internetwork (IN) pixels (blue and light blue, degrees of polarization less than three times the noise level) for the data on 2014 May 10 and those on 2019 August 10. Red represents the active region and network boundary pixels with the offsets of the peak positions of the Stokes $V/I$ signals at 6 steps or more from the line center; light red represents those offset 5 steps or less. Blue stands for the internetwork pixels with the offsets of the peak positions at 3 steps or less from the line center, and light blue stands for those offset at 4 steps or more. 
The disk was divided into four zones having different distances from the disk center for the analysis of the center-to-limb variation (CLV). Their borders are indicated by dashed lines. The inset shows the colors representing the four zones, which are used in later analysis.
  \label{fig:fig4}}
\end{figure}

\begin{figure}
\plotone{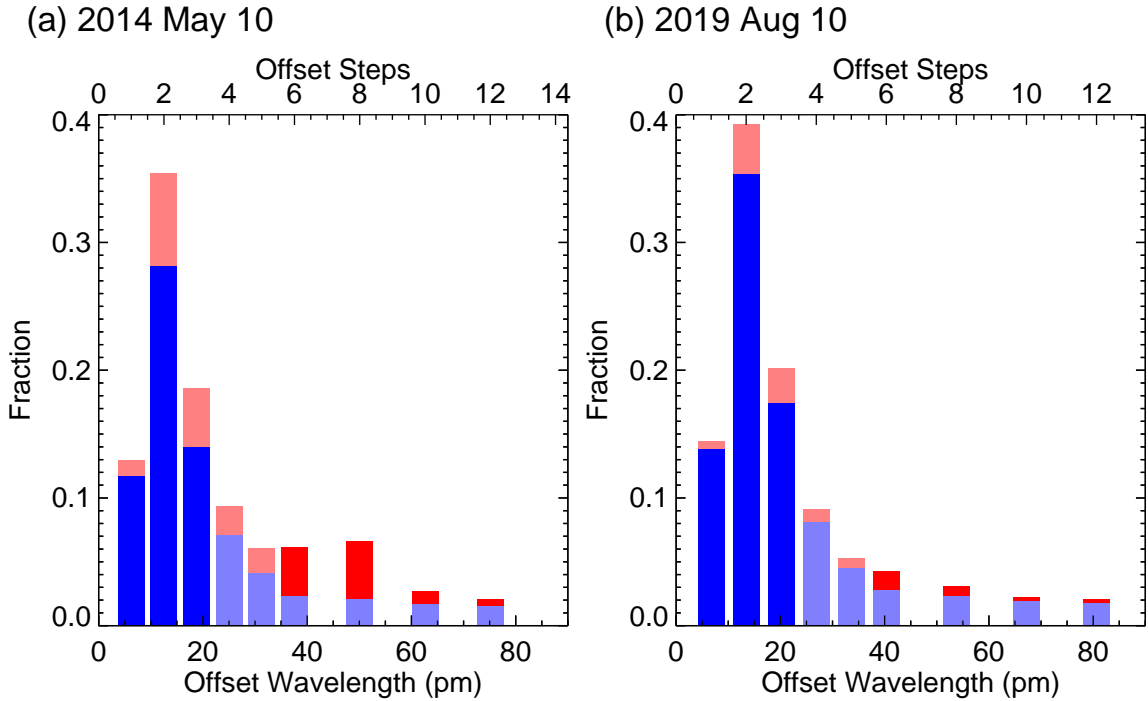}
\caption{Histogram of the wavelength offsets from the line center of the peak positions of the Stokes $V/I$ signals for the data on 2014 May 10 and those on 2019 August 10. The fractions of the internetwork regions (blue and light blue) and those of the active regions or network boundaries (red and light red) are stacked, and are shown with the same colors as those in Figure 4.
  \label{fig:fig5}}
\end{figure}

To study the properties of the Stokes $V/I$ signals of the internetwork fields more quantitatively, we isolated the internetwork pixels in the Stokes $V/I$ maps, excluding the pixels in the active regions or the network boundaries in the quiet Sun. 

Figure \ref{fig:fig3} suggests that the offsets of the peak positions of the Stokes $V/I$ signals could be a criterion to distinguish the active regions / network boundaries and the internetwork regions. However, as seen above, some of the pixels with large offsets do not correspond to the kG magnetic field. Furthermore, it is presumed that there are pixels including both the kG component and the hG one and they show high degrees of polarization in a wide range of the wavelengths. The peak wavelength offsets of such pixels are not necessarily as large as the ordinary kG pixels.
Therefore, it is important to take the degree of polarization into account. 
The wavelength offsets of 6--12 steps (37.7 / 40.3 -- 75.5 / 80.5 pm) correspond to the Zeeman splitting of the magnetic field of about 1--2 kG, and the pixels showing substantial polarization in this range can be considered to include kG magnetic elements inside. Then we defined the pixels containing the kG magnetic field by using the maximum absolute (unsigned) degree of polarization of Stokes $V/I$ signals at the offsets from the line center of 6, 8, 10, and 12 steps. If the polarization at any of these offsets exceeds three times of theRMS noise level, the corresponding pixel is presumed to include the kG magnetic field. The influence of the choice of the threshold to the results is mentioned in the next subsection.
With this criterion, we classified the pixels containing kG field as active regions or network boundaries in the quiet Sun, and the pixels free from kG components as internetwork pixels.

In Figure \ref{fig:fig4}, the defined active region and network boundary pixels are shown in red and light red for the data on 2014 May 10 and 2019 August 10. The threshold of the degree of polarization of the Stokes $V/I$ signals used to define the active region and network boundary pixels was 0.00048 (3$\times$ RMS noise level mentioned in Section 2.1) after the replacement of the polarization modulator in 2014. The pixels in red show the offsets of the peak positions of the Stokes $V/I$ signals at 6 steps or more from the line center, and the maximum degree of polarization of the Stokes $V/I$ signals exceeds 0.00048. 
On the other hand, the pixels in light red show the offset of the peak positions within 5 steps from the line center, yet they show the Stokes $V/I$ signals exceeding 0.00048 at the wavelength offset of 6 steps or more. 
The pixels in light red are distributed in the vicinity of the red pixels, and they correspond to the active regions and the network boundaries seen in Figures \ref{fig:fig2}(c) and \ref{fig:fig2}(d) as well as the pixels in red. This fact supports the classification of the pixels in light red as the active region or network boundary pixels. 

The remainder of the pixels (with degrees of polarization less than three times the noise level at the offset of 6 steps or further) are considered to be the internetwork pixels. We particularly focus on those which have the offsets of the peak positions of the Stokes $V/I$ signals within 3 steps from the line center; they are analyzed later and shown in blue in Figure \ref{fig:fig4}. The pixels in light blue have the peak offset at 4 steps or more from the line center and they scatter widely over the entire disk. As discussed above, the pixels with large peak offsets in the internetwork pixels are, in many cases, non-magnetic or cancelled-magnetic pixels.

Figures \ref{fig:fig5}(a) and \ref{fig:fig5}(b) show histograms of the wavelength offsets of the peak positions in the Stokes $V/I$ profiles for the data on 2014 May 10 and those on 2019 August 10. The fractions of the active region / network boundary pixels and those of the internetwork pixels are shown in the same colors as in Figure \ref{fig:fig4}. The fractions of the pixels in the active regions or network boundaries stand out at the offset of 6 or 8 steps in panel (a) owing to the high solar activity.
On the other hand, a conspicuous peak formed by the internetwork pixels can be found around the offset of 2 steps, namely 12.6 / 13.4 pm from the line center in both panels. These wavelength offsets correspond to the Zeeman splitting of 365 / 392 G. However, as stated in Section 2.2, the offset of 2 steps corresponds to the Stokes $V/I$ peak position of the weak field regime. Thus, while the definite magnetic field strength of the internetwork pixels is unknown, it is presumed to be 300--400 G or less.
Though the results of the former measurements of the internetwork magnetic field showed some scattering, they were mostly up to 350 G \citep{2003A&A...408.1115K}. Therefore, the result 300--400 G or less does not contradict the previous results.

The internetwork pixels and the active region or network boundary pixels in the quiet Sun are defined for all the 99 sets of data selected for this study. Owing to the replacement of the polarization modulator, the RMS noise level was much reduced after 2014, as mentioned in Section 2.1. The 3$\times$RMS noise level changed from 0.00138 to 0.00048. It is presumed that the higher threshold applied to data until 2013 to define the active region or network boundary pixels might have missed some of them, and it should be noted that in such cases some of them are mistakenly classified as the internetwork pixels.

\subsubsection{Center-to-Limb Variation of the Internetwork Stokes $V/I$ Signals}

\begin{figure}
\plotone{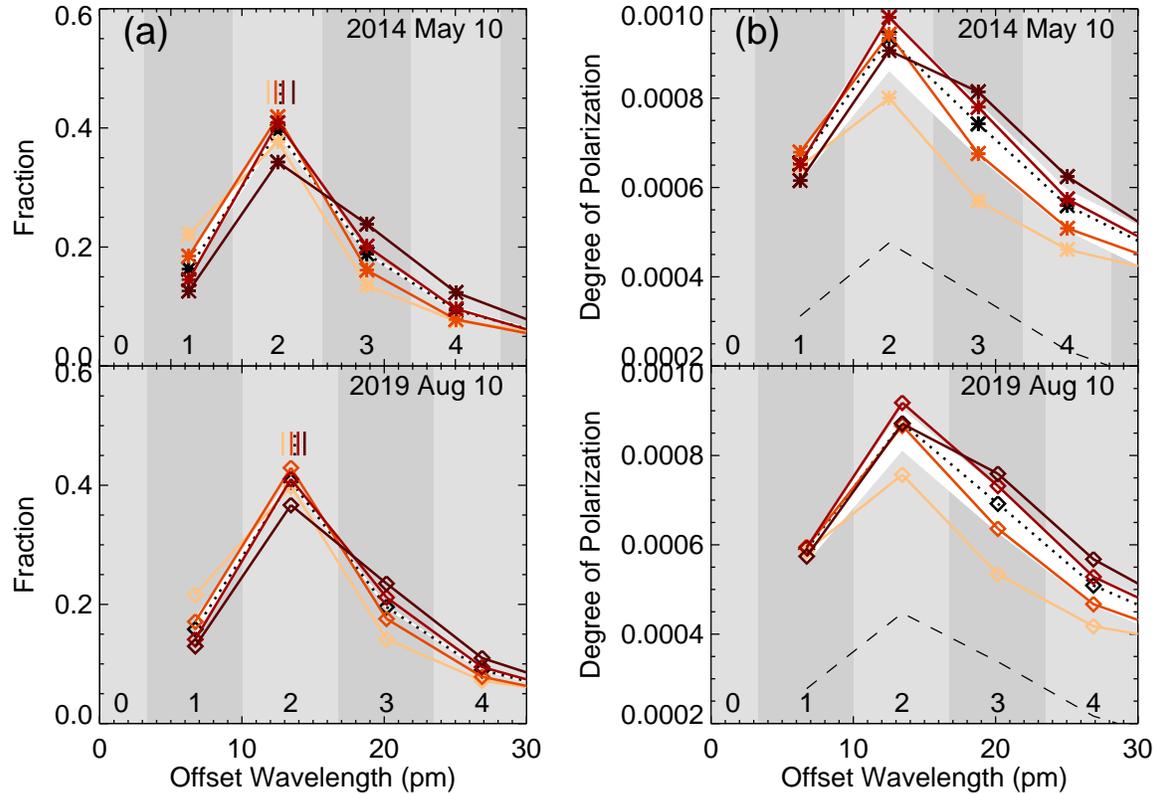}
\caption{
(a) Histograms of the peak position offsets from the line center of the Stokes $V/I$ signals of the \ion{Fe}{1} 1564.8 nm line for the pixels in the internetwork regions. Those on 2014 May 10 and 2019 August 10 are shown.
The black dotted lines indicate the histograms for the entire disk ($\theta < 75^{\circ}$), and the colored lines indicate those for four zones with different distances from the disk center, as shown in Figure \ref{fig:fig4}. The plausible maximum positions of the fraction distributions are indicated with the vertical bars. The gray stripes in the background show the wavelength sampling steps. The numbers of offset steps from the line center are labelled at the bottom of each panel. 
The results shown here are based on the kG-field threshold of 3$\times$RMS noise level, but in addition, the range of the histograms for the entire disk based on the thresholds of 2$\times$ and 4$\times$ RMS noise level are also shown with white strips around the dotted lines.
(b) Average degrees of the polarization for each peak position offset of the Stokes $V/I$ signals. The values for the entire disk and the four zones are shown in the same way as in panel (a). In addition, the standard deviations of the degrees of the polarization of the entire disk are shown with dashed lines.
As in panel (a), the range of the average degrees of the polarization for the entire disk based on the thresholds of 2$\times$ and 4$\times$ RMS noise level are shown with white strips around the dotted lines.
  \label{fig:fig6}}
\end{figure}

Next, we describe the detailed properties of the Stokes $V/I$ signals of the internetwork magnetic fields, and focus on their center-to-limb variation. We divided the solar disk into four zones, one circle and three annuli, which correspond to the angular distances $\theta$ from the disk center of $\theta < 20^{\circ}$, $20^{\circ} < \theta < 40^{\circ}$, $40^{\circ} < \theta < 60^{\circ}$, and $60^{\circ} < \theta < 75^{\circ}$, as shown in Figure \ref{fig:fig4}. The area beyond $75^{\circ}$ was not used for the analysis because the position of the limb of the Stokes images could not be accurately determined. In Figure \ref{fig:fig6} (and in later presentations), these zones are represented by different colors, from light brown to dark brown, as shown in the inset of Figure \ref{fig:fig4}.

Figure \ref{fig:fig6}(a) shows the histograms of the offsets from the line center of the peak position of the Stokes $V/I$ signals in the internetwork regions for the data from 2014 May 10 and 2019 August 10. The dotted lines indicate the histograms for the entire disk ($\theta <75^{\circ}$). As already shown in Figure \ref{fig:fig5}, the pixels with the offset of 2 steps were most numerous on both dates. This histograms are based on the threshold of 3$\times$RMS noise level to remove the kG-field as described in the previous subsection. To check the influence of the choice of the threshold, we calculated the histograms based on the thresholds from 2$\times$ to 4$\times$ RMS noise level, and their ranges are also shown with white strips around the dotted lines. The fact that the strips are very narrow means that the results hardly depend on the choice of the threshold.

Although the maximum of the fraction in each histogram falls at the offset of 2 steps, taking the fractions at the offsets of 1 and 3 steps into the consideration, we defined the plausible position of the ``true'' maximum as the vertex of a parabola fitted to the fractions at the offsets of 1, 2, and 3 steps. The offsets of the maximum positions thus derived for the whole disk for the two dates are indicated by vertical bars drawn with dotted lines.
The offset of the maximum position on 2019 August 10 was somewhat larger than that on 2014 May 10. However, the wavelength sampling widths on these dates were not identical, and the difference between these offsets was much smaller than the sampling widths. Therefore, it cannot be concluded that there was a difference between the distributions of the offsets on two observing dates near the most active period and the most quiet period of the Sun.

In Figure \ref{fig:fig6}(a), the light-brown lines indicate the histograms of the zone of $\theta < 20^{\circ}$, and the lines get darker toward the limb. All the zones also show that the fraction at the 2-steps offset was most numerous. However, the plausible ``true'' offsets from the line center of the maximum fraction indicated in Figure \ref{fig:fig6}(a) with colored vertical bars show a slight increase with the increase of the distance from the disk center. This is the common property for the data from both 2014 May 10 and 2019 August 10.

This does not necessarily mean that the magnetic field strength increases toward the limb and it leaves the weak field regime. The average Stokes $I$ profiles of the internetwork region in the four zones normalized at the far wing are shown in Figure \ref{fig:fig1}(a). The brightness in the wing decreases from the disk center to the limb. Figure \ref{fig:fig1}(b) shows the relative difference of the profiles with respect to that around the disk center. The brightness around 20 pm from the line center showed the most notable decrease, which means that the width of the \ion{Fe}{1} 1564.8 nm line increased from the disk center to the limb. The peak position in a Stokes $V/I$ profile in the weak-field regime is located where the gradient of the Stokes $I$ profile takes the largest value. The profiles shown in Figure \ref{fig:fig1} indicate that the maximum Stokes $V/I$ signal in the weak-field regime moved further from the line center with the increase of the angular distance from the disk center. Therefore, the systematic difference in the peak position seen in Figure \ref{fig:fig6}(a) was produced by the broadening of the \ion{Fe}{1} 1564.8 nm line toward the limb.

In Figure \ref{fig:fig6}(b), the average absolute (unsigned) degree of polarization of the pixels is shown at each peak position offset of the Stokes $V/I$ signals. While the offsets of the peak positions of the Stokes $V/I$ signals discussed above  just give the upper limit of the absolute magnetic field strengths in the weak field regime, the degree of polarization is proportional to the longitudinal (line-of-sight) component of the internetwork magnetic fields. The dotted lines show the average degrees of polarization for the entire disk ($\theta<75^{\circ}$), and the dashed lines show their standard deviations. The average degree reached its maximum at 2-steps offset from the line center, where the fractions shown in Figure \ref{fig:fig6}(a) also reached their maximum. 
The range of the average degrees of polarization based on the thresholds from 2$\times$ to 4$\times$ RMS noise level are also shown with white strips. The behavior of the average degrees of polarization does not depend on the choice of the threshold.

The degrees of polarization of the zone including the disk center ($\theta < 20^{\circ}$) shown with the light-brown lines in Figure \ref{fig:fig6}(b) were clearly lower than that of the other zones except at the offset of 1 step. Furthermore, there was a tendency for the degree of polarization to increase with the angular distance from the disk center. This can be considered an indication that the horizontal component dominates in the internetwork field and longitudinal magnetic field components observed as the Stokes $V/I$ signals were small near the disk center. 

At the offset of 2 steps in Figure \ref{fig:fig6}(b), it is evident that the degree of polarization increased until the $40^{\circ} < \theta < 60^{\circ}$ zone. However, the zone of $60^{\circ} < \theta < 75^{\circ}$ shows a lowering of the polarization. This does not necessarily contradict the assumption that the horizontal component dominates in the internetwork magnetic fields; we must consider the degradation of the spatial resolution toward the limb due to the foreshortening effect. In the internetwork region small positive and negative polarities are mixed within a small spatial scale. The foreshortening increases the chance that both the polarities are included within a pixel and cancel each other out. 
Thus, we can presume that the degree of polarization near the limb, particularly that in the $60^{\circ} < \theta < 75^{\circ}$ zone, suffers a lowering of the polarization.

The dominance of the horizontal field is sometimes considered to be an apparent effect appearing in isotropically distributed orientations, where the frequency of the inclination is proportional to sine of the inclination \citep{2011ASPC..437..451S}. If this is true, the Stokes $V/I$ signals seen in the full-disk maps show the same properties regardless of the angular distance from the disk center. However, the results show an increase of the degree of polarization toward the limb. In principle, this indicates the true dominance of the horizontal orientation in the internetwork magnetic field. 
Nevertheless, we should note the possibility that the degree of dominance of the horizontal field depends on the angular distance from the disk center, because the formation height of the \ion{Fe}{1} 1564.8 nm line depends on it.
In fact, \citet{2013A&A...555A.132S} noted that the inclination distribution depends on the height, and higher layers show more horizontal magnetic fields. This view is supported by \citet{2016A&A...593A..93D}.

\subsection{Stability of Various Properties of the Stokes $V/I$ Signals} \label{subsubsec:hide}

Next, we investigated the stability of the various properties of the Stokes $V/I$ signals of the internetwork magnetic fields seen in Section 3.1 during the years of 2010--2019. This period covers most of solar cycle 24, as seen in Figure \ref{fig:fig7}(a), where the sunspot number obtained by the NAOJ is presented. As mentioned in Section 2, we picked 99 data from the Fe 1564.8 nm line during this period.

\subsubsection{Wavelength Offsets of the Peak Absolute Polarization} \label{subsubsec:hide}

\begin{figure}
\plotone{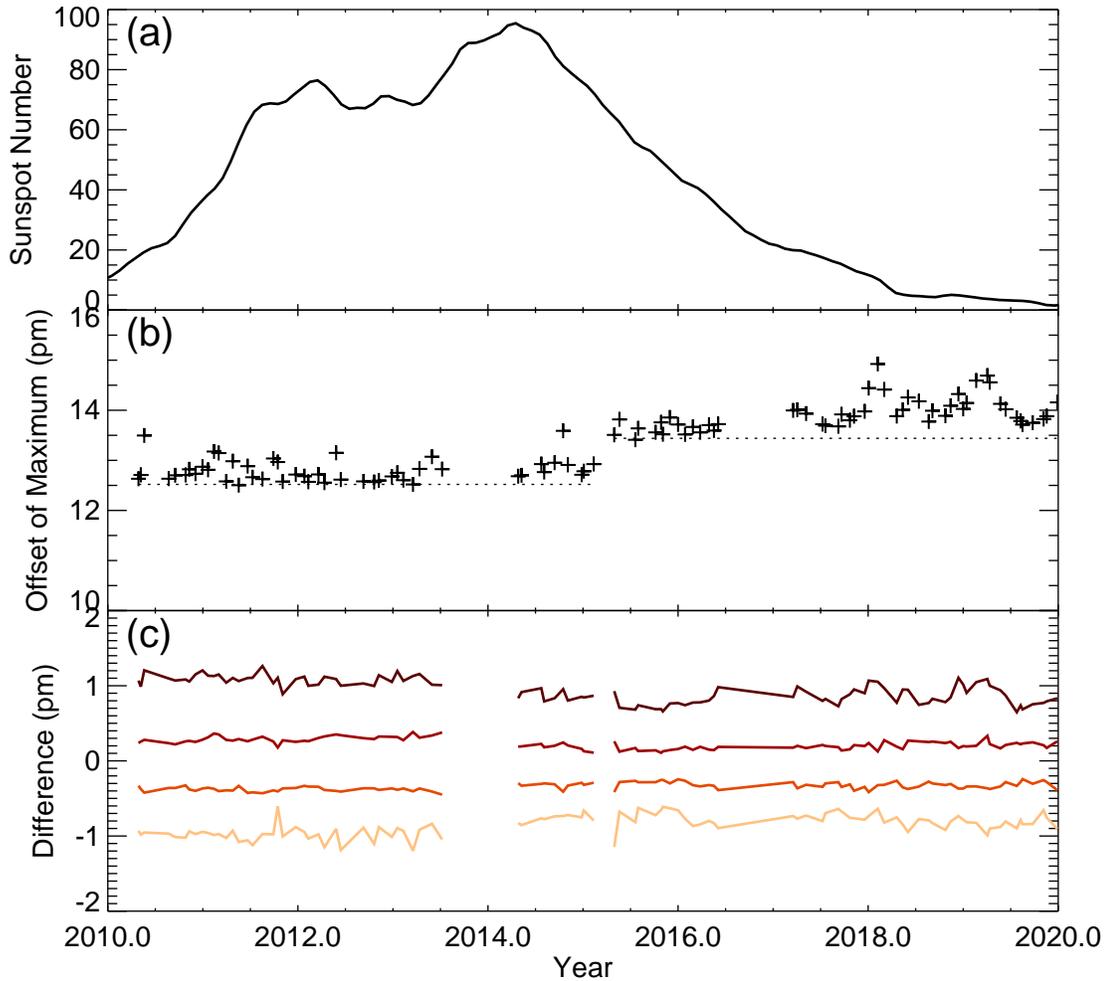}
\caption{Time variation of the sunspot number and those of the offsets of the fitted maxima of the distributions of the peak positions of the Stokes $V/I$ signals of the internetwork pixels from 2010--2019. Panel (a) shows the sunspot relative numbers obtained by the NAOJ. Panel (b) shows the variation of the offset of the fitted maximum calculated from the internetwork pixels in the entire disk ($\theta < 75^{\circ}$). The position of the offsets of 2 steps from the center of the \ion{Fe}{1} 1564.8 nm line is indicated by dotted lines; it was changed due to the modification of the optical system in 2015. Panel (c) shows the difference in the offsets of the maxima with respect to the values shown in panel (b) at various distances from the disk center. The colors of the lines correspond to the zones shown in the inset of Figure \ref{fig:fig4}.
  \label{fig:fig7}}
\end{figure}

Figure \ref{fig:fig7} shows the time variation of the offsets of the fitted maximum positions of the distributions of the peak positions of the Stokes $V/I$ signals, which are shown with vertical bars in Figure \ref{fig:fig6}(a) for the data from 2014 May 10 and 2019 August 10. The variation of the sunspot number is also presented. The fitted offsets for all internetwork pixels ($\theta < 75^{\circ}$) are shown with plus signs in Figure \ref{fig:fig7}(b). The long gap extending over 2013--2014 was due to a failure caused by a lightning strike. After this gap, the polarization modulator was changed and the noise level was greatly improved. The long gap from 2016--2017 was due to a repair work. During the short gap in 2015, the optical system was replaced and the wavelength sampling step was changed from 6.3 pm to 6.7 pm. In Figure \ref{fig:fig7}(b), the offset wavelengths corresponding to 2 steps from the line center are shown with dotted lines. 
As mentioned in Section 3.1, the offsets of the fitted maxima of the distributions of the Stokes $V/I$ peak positions show a systematic difference across the gap in 2015 due to the change of the spectral sampling, which is a kind of quantization error due to coarse sampling. The offset remained slightly above the offset of 2 steps through the entire period. Therefore, in spite of the systematic differences, we can conclude that there was no detectable variation tracking the solar activity in the offsets of the fitted maxima, while the sunspot number showed conspicuous variation.

Figure \ref{fig:fig7}(c) shows the variations of the relative wavelength offsets of the four zones with different distances from the disk center with respect to the offset of all pixels shown in Figure \ref{fig:fig7}(b). The colors of the lines correspond to the zones shown in the inset of Figure \ref{fig:fig4}. Figure \ref{fig:fig6} shows that the wavelength offset increased from the disk center to the limb. We can confirm that this tendency was consistent from 2010--2019. Until 2013, the differences among the four areas were larger than those after 2014, likely due to the differences in the noise level. 
Therefore, we can conclude that there was no detectable cycle variation in the wavelength offset of the maximum of the distribution of the peak positions of the Stokes $V/I$ signals in the internetwork fields. This means that the strength of the internetwork magnetic fields remain within the weak field regime throughout a solar cycle.

\subsubsection{Average Degrees of Polarization} \label{subsubsec:hide}

\begin{figure}
\plotone{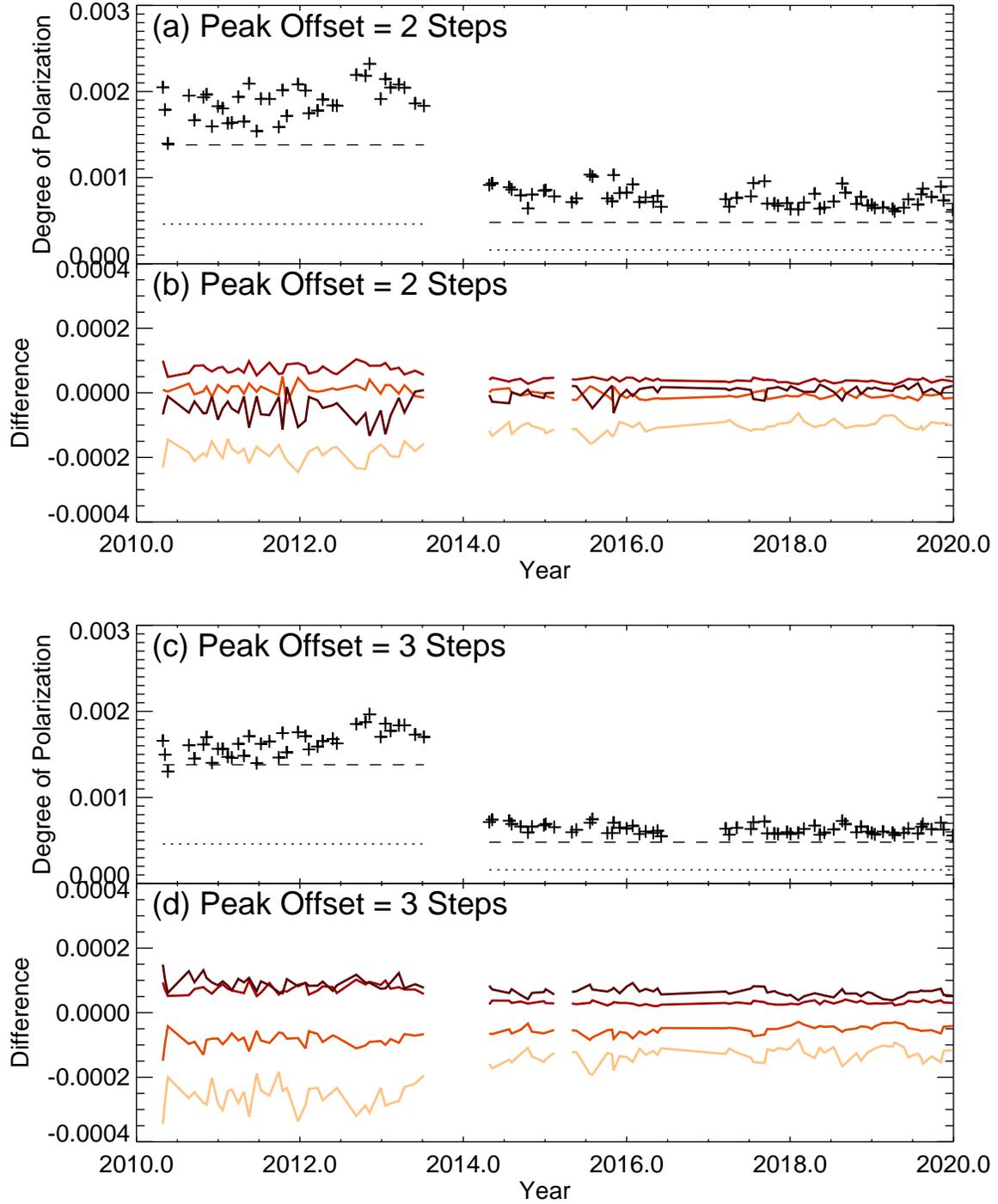}
\caption{Time variations of the average degrees of the polarization of the internetwork pixels. Panels (a) and (b) show the polarization of the pixels where the offset of the peak positions of the Stokes $V/I$ signals is 2 steps from the line center, and panels (c) and (d) show that for 3 steps. Panels (a) and (c) show the polarizations calculated from the internetwork pixels in the entire disk ($\theta < 75^{\circ}$). The degrees of polarization corresponding to 1$\times$ and 3$\times$ RMS noise level are shown with dashed and dotted lines, respectively; these levels jumped due to the replacement of the polarization modulator in 2014. Panels (b) and (d) show the differences in the degree of polarization from the values in panels (a) and (c) at various distances from the disk center. The colors of the lines correspond to the zones shown in the inset of Figure \ref{fig:fig4}.
  \label{fig:fig8}}
\end{figure}

Figure \ref{fig:fig8} shows the time variations of the average degrees of polarization of the pixels where the offset of the peak positions of the Stokes $V/I$ signals is 2 or 3 steps from the line center.
The degrees of polarization calculated from the entire internetwork pixels ($\theta < 75^{\circ}$) for the offset of 2 steps are shown in Figure \ref{fig:fig8}(a) and those for 3 steps are shown in Figure \ref{fig:fig8}(c). The values for the data from 2014 May 10 and 2019 August 10 have already shown in Figure \ref{fig:fig6}(b). 
The $1\times$ and $3\times$ RMS noise levels in the Stokes $V/I$ signals both before and after the gap of 2013--2014 are shown with dotted lines and dashed lines. The average degree of polarization until 2013 and that from 2014 did not show systematic change during their respective periods. Figures \ref{fig:fig8}(b) and \ref{fig:fig8}(d) show the relative variations of the average degrees of polarization in the four zones with different angular distances from the disk center with respect to the degrees of polarization of all pixels shown in Figures \ref{fig:fig8}(a) and \ref{fig:fig8}(c). The colors of the lines correspond to the zones shown in the inset of Figure \ref{fig:fig4}. As mentioned in Section 3.1.4, Figure \ref{fig:fig6} shows that the average degree of polarization at the offset of 2 steps increases until the $40^{\circ} < \theta < 60^{\circ}$ zone toward the limb and that the lowering of the polarization is seen in the $60^{\circ} < \theta < 75^{\circ}$ zone. Figure \ref{fig:fig8}(b) shows a similar tendency throughout the measured period. The relative differences until 2013 and those after 2014 showed a systematic difference, which was likely due to the changes in noise levels. However, the general characteristics were maintained during the respective periods. 
The relative differences of the average degrees of polarization at the offset of 3 steps in Figure \ref{fig:fig8}(d) show increases of the polarization toward the limb throughout the measured period; this tendency became clearer after 2014 than it was before 2013. Nonetheless, these relative differences show no detectable systematic variations until 2013 or after 2014.

However, it should be noted that a possible slight increase in the average polarization until 2013 was evident in Figures \ref{fig:fig8}(a) and \ref{fig:fig8}(c). This is presumably due to contamination from the increasing non-internetwork pixels, because the thresholds for determining the active region or network boundary pixels were high until 2013. Some of the pixels with a polarization slightly below the threshold, which would be classified as active regions or network pixels with the threshold after 2014, were classified as internetwork pixels until 2013.

From the above, we can confirm that there was no obvious change in the properties of the degree of polarization throughout the period of 2010--2019. As a whole, there is no result showing changes correlating with the solar activity variation. Thus, we can conclude that the magnetic field properties of the internetwork regions show no detectable cycle variation.

\section{Summary} \label{sec:summary}

In this paper, we presented the properties of the internetwork magnetic field in the solar photosphere derived from the full-disk \ion{Fe}{1} 1564.8 nm polarization data, which were obtained with the spectropolarimeter of the SFT at the NAOJ. The internetwork field has been studied by many researchers; however, our method is quite different from most of the previous studies. Instead of large, advanced telescopes, we used a synoptic instrument with moderate spatial and spectral resolutions, and a field of view sufficiently to cover the full Sun. In addition, we used data regularly obtained during 2010--2019 for the analysis, opposed to the {\it ad hoc} observations with a small field of view used in previous studies. In those studies, the method of analysis was mostly based on the inversions of the polarization signals; however, we statistically analyzed large amount of polarization data without inversions. Although we were not able to obtain various parameter values from the Stokes inversions, the results were robust and free from the complexities arising from inversions. Thus, the analysis here shows the properties of the internetwork field derived from a quite different point of view from those in the previous studies.

Nevertheless, the obtained results showed that the internetwork is filled with small-scale magnetic fields, the strengths of which are within the weak-field regime of the \ion{Fe}{1} 1564.8 nm line (300--400 G or less). Moreover, the results showed the center-to-limb variation of the polarization signal, which is consistent with the former studies that indicated the horizontal orientation of the internetwork fields. These results are consistent with those of the majority of previous studies. There have been some discrepancies in the previous results. In particular, the results based on the observations of the \ion{Fe}{1} 1.56 $\mu$m lines often show inconsistencies; however, our results based on the analysis of the data taken with the \ion{Fe}{1} 1564.8 nm line support the majority of the previous results.

We analyzed only the Stokes $V/I$ signals here, but the data contain full-Stokes information. Therefore, to expand the study using the linear polarization signals is the next step. Furthermore, in addition to the \ion{Fe}{1} 1564.8 nm line, our data include the Stokes signals of the \ion{Fe}{1} 1565.3 nm line, which is close to the \ion{Fe}{1} 1564.8 nm line and also shows remarkable Zeeman splitting. Comparison between the results presented here and those based on the \ion{Fe}{1} 1565.3 nm line will be a subject for future work.

Furthermore, by analyzing the data from 2010--2019 covering most of solar cycle 24, we confirmed that the properties of internetwork fields do not show notable cycle variation. 

It has been inferred that the weak horizontal magnetic field of the solar internetwork regions is only detectable with observations of high spatial resolutions (\citeauthor{2009A&A...502..969B}, \citeyear{2009A&A...502..969B}; see also the discussion in \citeauthor{2019LRSP...16....1B}, \citeyear{2019LRSP...16....1B}).
However, this study shows that the internetwork magnetic field can be seen even if the spatial resolution is not very high. Nevertheless, we are aware of that our observations show only limited aspects of the internetwork field, and it was proved that the quantitative results are strongly influenced by the noise levels (Figure \ref{fig:fig8}) and wavelength sampling (Figures \ref{fig:fig6} and \ref{fig:fig7}). 
In addition, we observed that the data with a high scattered light level obtained during a modification of the instrument (not used in the analysis) showed an apparent increase in the wavelength offset from the line center of the peak positions of the Stokes $V/I$ signals. 
These facts imply that we must be careful when comparing observational results of the internetwork fields obtained with different instruments and pay close attention to the observation conditions to perform a quantitative comparison of the results obtained under different conditions.

\acknowledgments

This work is supported by JSPS KAKENHI Grant Number JP17204014 and JP23244035, and by the NAOJ research grant.
We appreciate the staff members of the Solar Science Observatory who have been involved in the daily operation of the instruments and the maintenance of the data servers. The authors are grateful to the anonymous referee for his/her careful reading and helpful comments. We would like to thank Editage (www.editage.com) for English language editing.


\end{document}